%% file: casares_v4.tex
\documentclass[graybox]{svmult}

\usepackage{mathptmx}       % selects Times Roman as basic font
\usepackage{helvet}         % selects Helvetica as sans-serif font
\usepackage{courier}        % selects Courier as typewriter font
\usepackage{type1cm}        % activate if the above 3 fonts are
\usepackage{makeidx}         % allows index generation
\usepackage{graphicx}        % standard LaTeX graphics tool
\usepackage{multicol}        % used for the two-column index
\usepackage[bottom]{footmisc}% places footnotes at page bottom

\makeindex             % used for the subject index
\begin{document}

\title*{New Optical Results on $\gamma$-ray Binaries}
\author{J. Casares, J.M. Corral-Santana, A. Herrero, J.C. Morales, T.
Mu\~noz-Darias, I. Negueruela, J.M. Paredes, I. Ribas, M. Rib\'o, D. Steeghs, 
L. van Spaandonk \and F. Vilardell}
\institute{J. Casares \at Instituto de Astrof\'\i{}sica de Canarias (IAC), 
E-38200 La Laguna, Tenerife, Spain and Departamento de Astrof\'\i{}sica, 
Universidad de La Laguna (ULL), E-38205 La Laguna, Tenerife, Spain,  
\email{jcv@iac.es}
\and $\star$ The affiliation of other co-authors is listed in the
acknowledgements.} 
%\and A. Herrero \at Instituto de Astrof\'\i{}sica de Canarias (IAC), 
%E-38200 La Laguna, Tenerife, Spain and Departamento de Astrof\'\i{}sica, 
%Universidad de La Laguna (ULL), E-38205 La Laguna, Tenerife, Spain , 
%\email{ahd@iac.es}, 
%\and J.C. Morales \at Institut d'Estudis Espacials de Catalunya (IEEC), 
%Edif. Nexus, C/Gran Capit\`a, 2-4, 08034 Barcelona, Spain, \email{jcmorales@am.ub.es}, 
%\and I. Negueruela \at Departamento de F\'\i{}sica, Ingenier\'\i{}a de Sistemas y 
%Teor\'\i{}a de la Se\~nal, Universidad de Alicante, Apdo. 99, 03080 Alicante, 
%Spain, \email{ignacio.negueruela@ua.es}, 
%\and J.M. Paredes \at Departament d'Astronomia i Meteorologia and Institut de 
%Ci\`encies del Cosmos (ICC), Universitat de Barcelona (UB/IEEC), Mart\'\i{} 
%Franqu\`es 1, 08028 Barcelona, Spain, \email{jmparedes@ub.edu}, 
%\and I. Ribas \at Institut de Ci\`encies de l'Espai (CSIC-IEEC), Campus UAB, 
%Facultat de Ci\`encies, Torre C5, parell, 2a pl., E-08193 Bellaterra, Spain,
%\email{iribas@ice.csic.es} 
%\and M. Rib\'o \at Institut de Ci\`encies de l'Espai (CSIC-IEEC), Campus UAB, 
%Facultat de Ci\`encies, Torre C5, parell, 2a pl., E-08193 Bellaterra, Spain, 
%\email{mribo@am.ub.es}, 
%\and F. Vilardell \at Departamento de F\'\i{}sica, Ingenier\'\i{}a de Sistemas y 
%Teor\'\i{}a de la Se\~nal, Universidad de Alicante, Apdo. 99, 03080 Alicante, 
%Spain,\email{francesc.vilardell@ua.es}}
\maketitle

%\abstract*{}

\abstract{We present new optical spectroscopy of the $\gamma$-ray binary 
LS 5039. Our data show evidence for sub-orbital modulation in the radial 
velocities with amplitude $\sim$7 km/s and period $\sim P_{\rm orb}/4$. 
This short-term oscillation is stable over at least 7 years and it 
is likely triggered by non-radial oscillations of the O6.5V optical star. 
We also present the 
results of a spectroscopic campaign on MWC 148, the optical counterpart of 
the new $\gamma$-ray binary candidate HESS J0632+067. Long-term variations in 
the H$_{\alpha}$ and H$_{\beta}$ emission line parameters 
are clearly detected which, if modulated with the binary orbit, would imply a 
period $>$200 days.}

\section{Introduction}
\label{sec:1}

There are only three confirmed galactic High Mass X-ray Binaries (HMXBs) with 
persistent TeV emission (PSR B1259-63, LS 5039 and LSI +61 303, 
see~\cite{holder09}) and one recently proposed candidate (HESS J0632+067~\cite{hinton09}). The VHE 
emission in the 3 confirmed binaries is strongly modulated with the orbital 
period, suggesting that the emitter is rather compact and 
close to the massive star. However, the origin of the VHE emission remains 
unclear, with competing leptonic and hadronic scenarios which invoke either 
inverse Compton scattering of stellar photons or proton-proton collisions in 
a jet/pulsar wind scenario. 
Obviously, the nature of the compact star is an important ingredient in the 
different $\gamma$-ray production models. PSR B1259-63 contains a pulsar 
whereas a black hole is not ruled out in LS 5039 and LSI +61 303.  
In this context optical studies can bring new insights through constraining 
the compact object mass with dynamical studies.  

\section{Revised Orbital Solution in LS 5039}
\label{sec:2}

LS 5039 was the first $\gamma$-ray binary detected by HESS, with a TeV 
luminosity of $\sim$10$^{34}$ erg s$^{-1}$. It was previously classified as 
a HMXB with spatially resolved radio emission at milliarcsec scale, atributed 
to either a relativistic jet or a pulsar wind. 
The first determination of the 3.9d orbital period 
and system parameters was reported by \cite{casa05}. The orbital parameters 
were found to depend on the spectral lines used in the cross-correlation 
analysis. In particular, the Balmer and HeI solutions are blueshifted with
respect to the HeII solution and also show a larger scatter. This is expected 
if low excitation lines are contaminated by P-Cyg profiles and, hence, the HeII 
velocities provide a much better description of the 
orbit of the optical star~\cite{casa05}. A tentative low inclination angle in 
the range 23--27$^{\circ}$ was derived under the assumption that the 
companion star is pseudo-synchronized, i.e., synchronized at periastron. 
This results in a compact object of 3--5 M$_{\odot}$ and, thus, the possibility
that LS 5039 harbours a low-mass black hole (see \cite{casa05} for details).  
%, although this claim should be taken with caution given the several assumptions
%involved (see \cite{casa05} for details).  

\begin{figure}[ht]
\begin{center}
\begin{picture}(250,150)(50,25)
\put(0,0){\includegraphics{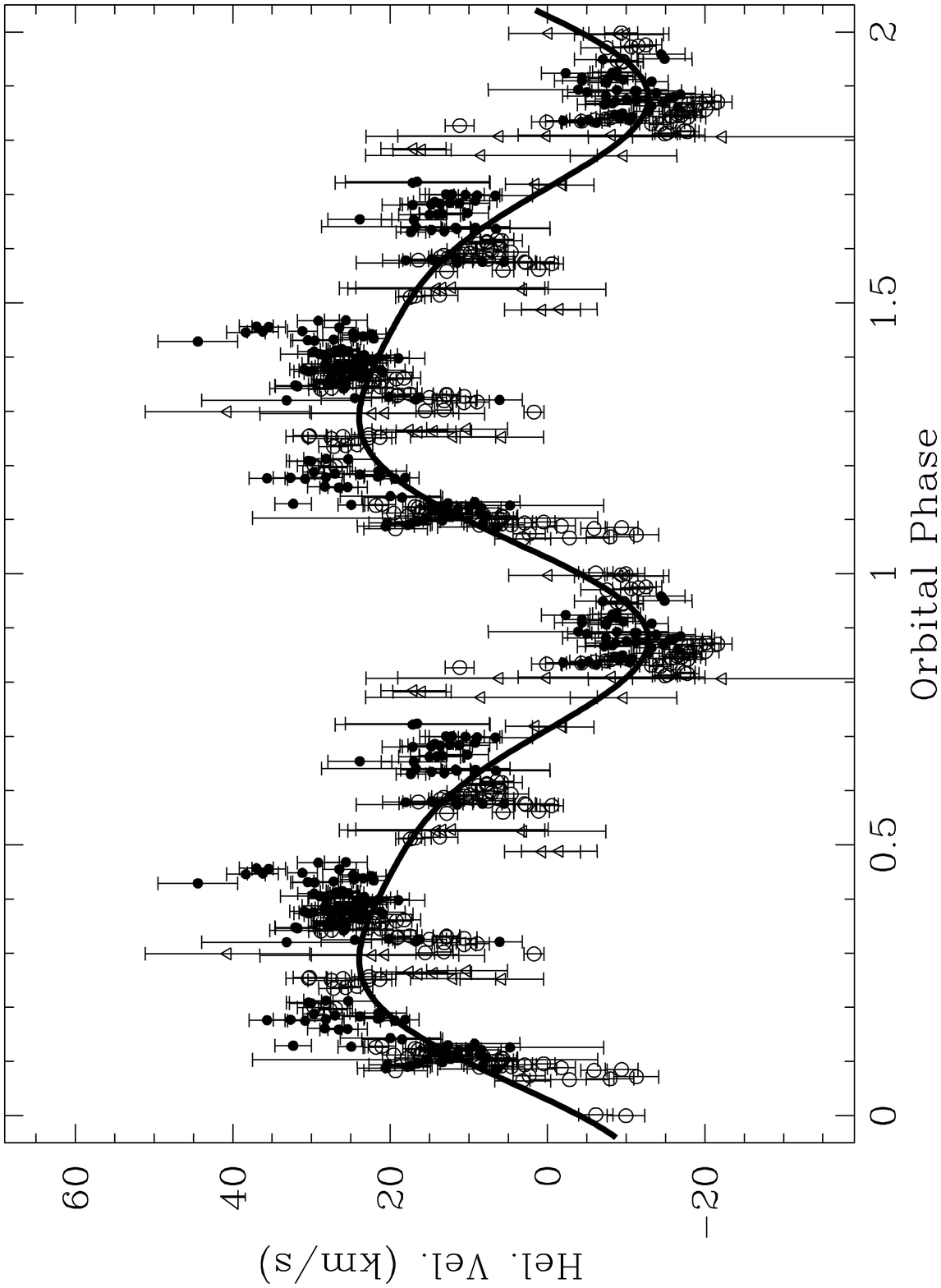}}
\put(0,0){\includegraphics{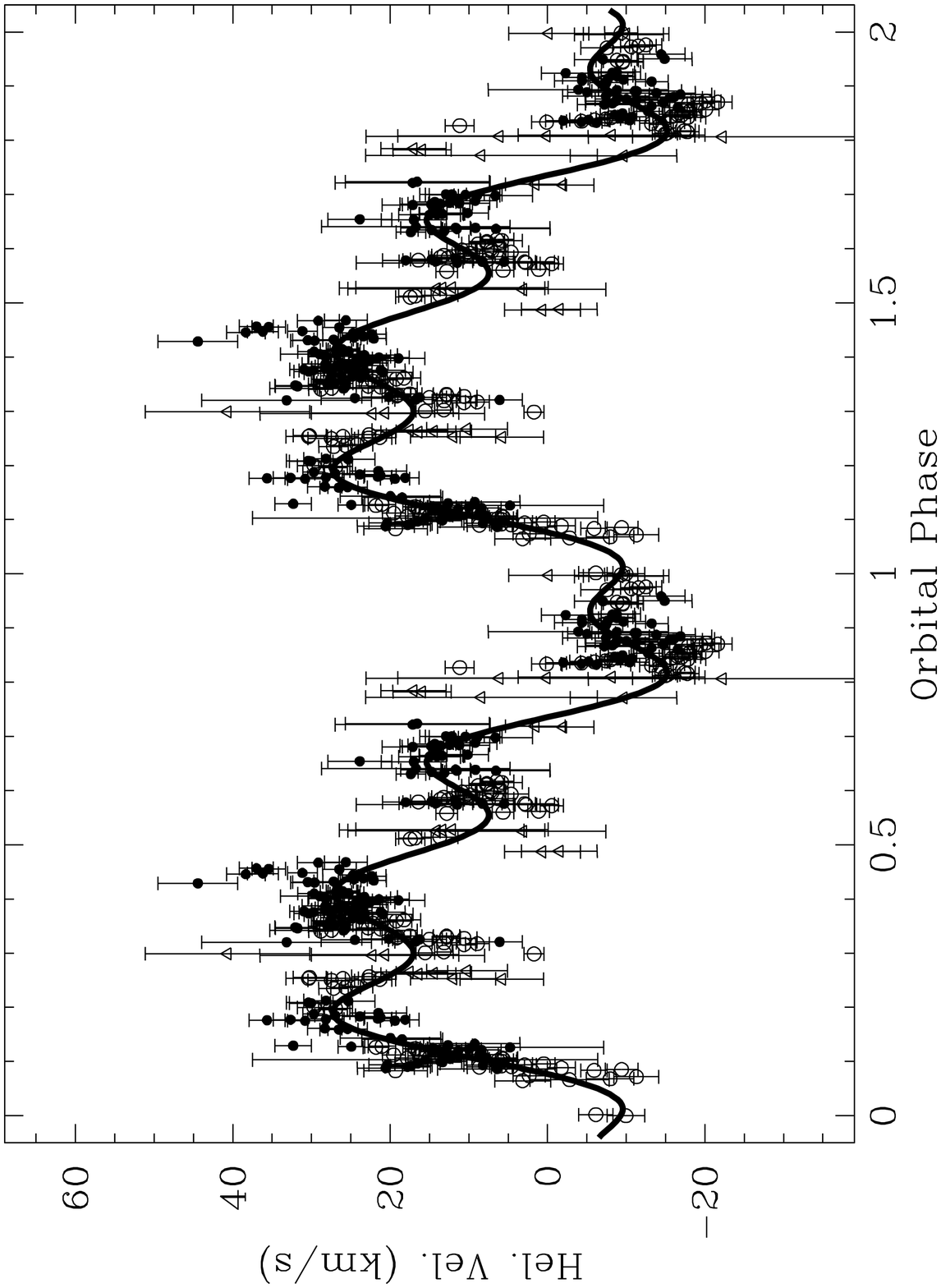}}
\noindent
\end{picture}
\end{center}
{\caption{Updated radial velocity curve in LS 5039. Left: velocities folded on
the best ephemeris with a two Fourier term fit sumperimposed. Filled circles
indicate INT data from 2002 \& 2003 reported in \cite{casa05}, open triangles
are SAAO data from 2005 and open circles INT data from 2007 and 2009. 
Right: Same as left but with an extra sine wave to account for the short-term
modulation.}} 
\label{fig:1}
\end{figure}

New spectrocopy was obtained in 2005, 2007 and 2009 using the 1.9m 
telescope at SAAO and the INT at La Palma. Fig.~1  
displays the new radial velocity points, together with data from \cite{casa05}, 
folded on the best orbital solution. The orbital modulation is clearly 
distorted by a short-term oscillation with amplitude $\sim$ 7 km s$^{-1}$ and 
a $\sim$1d period or $P_{\rm orb}/4$. In order to test the  
significance of this perturbation we fitted the eccentric orbital modulation  
with a 2-term Fourier series and obtain a reduced $\chi^2$ of 11.3 (left panel
in Fig.~1). We subsequently added an extra sine wave to account for the short 
period oscillation and find that the reduced $\chi^2$ drops to 7.1 (right 
panel in Fig.~1), thus confirming that the short-term modulation is 
highly significant. We note that this modulation must be stable over 
our 7 years database, otherwise it would be blurred in the phase folded radial 
velocity curve.  

\begin{table}
\caption{Orbital solutions for LS 5039}
\label{tab:1}       % Give a unique label
%\begin{tabular}{lcc}
\begin{tabular}{p{4.5cm}p{3.5cm}p{3.5cm}}
\hline\noalign{\smallskip}
Parameter & Eccentric Fit & Eccentric Fit\\
& & + 1d oscillation$^a$ \\
\noalign{\smallskip}\svhline\noalign{\smallskip}
$P_{\rm orb}$ (days)      & 3.90597$\pm$ 0.00009 & 3.90608$\pm$ 0.00008 \\
$T_0$ (HJD$-$2\,450\,000) & 2478.11$\pm$ 0.08 & 2478.08$\pm$ 0.06 \\
$e$                       & 0.30$\pm$ 0.03  & 0.35$\pm$ 0.03 \\
$w$ ($^{\circ}$)               & 215$\pm$ 6 & 212$\pm$ 5   \\
$\gamma$ (km~s$^{-1}$)    & 17.3$\pm$ 0.5  & 17.3$\pm$ 0.5  \\
$K_1$ (km~s$^{-1}$)       & 23.1$\pm$ 0.9 &  24.2$\pm$ 0.9 \\
$a_1 \sin i$ (R$_\odot$)  & 1.70$\pm$ 0.07  & 1.75$\pm$ 0.07  \\
$f(M)$ (M$_\odot$)        & 0.0043$\pm$ 0.0005 & 0.0047$\pm$ 0.0006 \\
rms of fit (km~s$^{-1}$)  &  8.3 &  6.4\\
\noalign{\smallskip}\hline\noalign{\smallskip}
\end{tabular}
$^a$ The period has been fixed to $P_{\rm orb}$/4
\end{table}

Ignoring this perturbation can certaintly bias the final orbital  
solution as illustrated in Table 1. 
The 2 columns compare the binary parameters obtained through a proper 
eccentric model fit to the HeII velocities with and without the 1d 
oscillation. Both, the eccentricity and radial velocity semi-amplitude 
$K_1$ are underestimated if the 1d oscillation is neglected. 
Also the periastron angle and the absolute phasing differ. The key binary
parameters are still consistent within 1-$\sigma$ but the difference can be 
larger when the fit is performed over 
scarcely sampled radial velocity curves. We note that this short-term 
modullation is very reminiscent of the  5 km s$^{-1}$ oscillation detected 
in Vela X-1 at 1/4 of the orbital period \cite{quaint03}. 
Following the work on Vela X-1 we propose that the 1d modulation in LS 5039 
is caused by tidally excited non-radial oscillations of the O6.5V star in
its eccentric orbit. 
The presence of this oscillation may also have 
an impact on the VHE radiation models because of the role played by the 
stellar photons and hence it should be investigated.

\section{Probing Binarity in MWC 148}
\label{sec:3}

MWC 148 is the optical counterpart of the point-like TeV source HESS J0632+067 
and it has been recently proposed as a candidate $\gamma$-ray binary based on 
the spectral properties and variability observed in both radio and 
X-rays~\cite{hinton09}.
In Oct 2008 we started a spectroscopic campaign to observe MWC 148 using 
several telescopes at the Teide and Roque de los Muchachos observatories. 
The averaged spectrum shows H$_{\alpha}$ and H$_{\beta}$ emission lines 
superimposed on a B0-type stellar spectrum dominated by HeI absorption lines 
(without any emission component). From the HeI profiles 
we estimate a rotational broadening $v \sin i= 370$ km/s. Since Be stars are 
known to rotate at typically 0.7--0.8 times the critical 
rotation velocity (i.e. $\sim$565 km/s for a B0V star) the binary inclination 
must be moderately high, in the range $\sim$55--70$^{\circ}$. 
Cross-correlation of the absorption lines with a broadened template star gives 
velocity excursions of a few tens km s$^{-1}$ but no clear periodicity is observed.

\begin{figure}[ht]
\sidecaption[t]
\put(0,0){\includegraphics{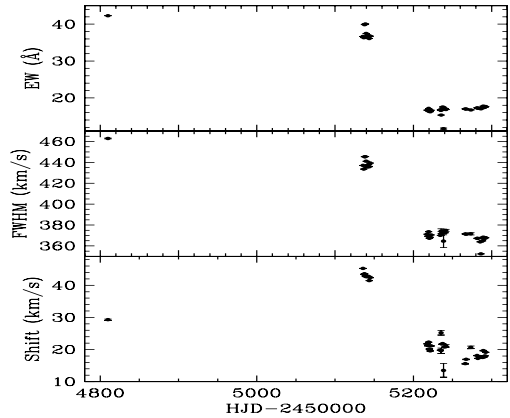}}
\parbox[t]{0.45\textwidth}
\caption{Time evolution of H$_{\alpha}$ emission line parameters in 
MWC 148 from October 2008 until April 2010. Top panel shows the EW, middle 
panel the FWHM and bottom panel the centroid velocity obtained through a 
Gaussian fit. The same trend is observed in H$_{\beta}$ but with lower 
amplitude.}
\label{fig:2}       % Give a unique label
\end{figure}

\vspace{2.0cm}

However, a dramatic change in the emission line parameters  
is detected between November 2009 and January 2010 (see Fig.~2). 
The lines become narrower, the EW drops by a factor 2 and radial velocity 
variations are also seen. If these changes are modulated with  
the binary phase then the orbital period must be longer than 200 days. However, 
superorbital $H\alpha$ variations due to precession of the circumstellar disc 
are also commonly seen in Be X-ray binaries~\cite{neg98}. Clearly many more 
observations are needed before the 
%binary parameters and, in particular, the 
orbital period and binary parameters can be constrained.    

\begin{acknowledgement}
Affiliation of co-authors is as follows: 
J.M. Corral-Santana and A. Herrero (IAC $\&$ ULL, Spain), J.C. Morales and 
I. Ribas (IEEC, Spain), T. Mu\~noz-Darias (Obs. Brera, Italy), I. Negueruela 
and  F. Vilardell (Univ. Alicante, Spain), J.M. Paredes and M. Rib\'o (Univ. 
Barcelona, Spain), D. Steeghs and L.van Spaandonk (Univ. Warwick, UK).  
J.C. acknowledges support from the Spanish MCYT through the project 
AYA2007-66887 and I.N. through project AYA2008-06166-C03-03. J.M.P. 
and M.R. acknowledge support by DGI of the Spanish MEC under grant 
AYA2007-68034-C03-01 and FEDER funds. M.R. also acknowledges 
support from MEC and European Social Funds through a 
\emph{Ram\'on y Cajal} fellowship.
Partly funded by the Spanish MEC under the Consolider-Ingenio 
2010 Program grant CSD2006-00070: first science with the GTC.  
\end{acknowledgement}
%

\input{references_casares.tex}

\end{document}

%% file: references_casares.tex
%%%%%%%%%%%%%%%%%%%%%%%% referenc.tex %%%%%%%%%%%%%%%%%%%%%%%%%%%%%%
% sample references
% %
% Use this file as a template for your own input.
%
%%%%%%%%%%%%%%%%%%%%%%%% Springer-Verlag %%%%%%%%%%%%%%%%%%%%%%%%%%
%
% BibTeX users please use
% \bibliographystyle{}
% \bibliography{}
%
\biblstarthook{}